\documentclass[11pt, letterpaper]{article}
\usepackage[utf8]{inputenc}
\usepackage[margin=1.5cm]{geometry}
\usepackage{titlesec}
\usepackage{tabularx}
\usepackage{array}
\usepackage{enumitem}
\usepackage{amssymb}
\usepackage{siunitx}
\usepackage{xcolor}
\newlist{selectlist}{itemize}{2}
\setlist[selectlist]{label=$\square$,leftmargin=*,noitemsep,topsep=0pt}
\usepackage{lmodern}
\usepackage[english]{babel}
\usepackage[numbers]{natbib}
\usepackage{subcaption}
\usepackage[colorlinks=true, citecolor=black, linkcolor=black, urlcolor=black]{hyperref}
\hypersetup{
    colorlinks=true,
    linkcolor=blue,
    filecolor=magenta,      
    urlcolor=blue,
}
\usepackage{graphicx}
\usepackage{csquotes}

\titleformat{\section}[block]{\hspace{1em}\bfseries}{\thesection.}{0.5em}{} 
\titleformat{\subsection}[block]{\hspace{1em}}{\thesubsection}{0.5em}{}

\begin{document}

\setlength{\parindent}{0pt}
\setlength{\parskip}{10pt}

\textbf{Article title}
\\{StimulHeat: a Low-Energy Wearable Thermal Feedback Device Using Peltier Elements with Heat Flow Controlled Loop for Hand Interactions in Virtual Reality
}\\

\textbf{Authors}
\\{Matthieu Mesnage$^1$, Sophie Villenave$^2$, Bertrand Massot$^1$$^*$, Matthieu Blanchard$^2$, Pierre Raimbaud$^2$, Guillaume Lavoué$^2$, Claudine Gehin$^1$
}\\

\textbf{Affiliations}
\\{$^1$ INSA Lyon, Ecole Centrale de Lyon, CNRS, Université Claude Bernard Lyon, CPE Lyon, INL, UMR5270, 69100 Villeurbanne, France
\newline $^2$ Ecole Centrale de Lyon, CNRS, INSA Lyon, Université Claude Bernard Lyon 1, LIRIS, UMR5025, ENISE, 42023 Saint Etienne, France
}\\

\textbf{Corresponding author’s email address}
\\{
bertrand.massot@insa-lyon.fr
}\\

\textbf{Abstract}
\\{Nowadays, the majority of wearable thermal feedback systems designed for use in virtual reality applications are not compatible or not integrated to standard controllers and are based on temperature control. The objectives of the present work is to enable integration with existing controllers, in this case Valve Index controllers, and to propose an alternative approach to managing thermal stimulation with Peltier modules by controlling heat flow instead of temperature. We introduce StimulHeat as a wireless, low power thermal feedback system, based on the continuous relationship between heat and current injection in thermoelectric device (TED). First, we designed an optimized TED driver capable of injecting a continuous, bidirectional current into the TED, thereby driving it as a heater or cooler. Subsequently, this driver was implemented in an electronic board to include temperature and heat flow control loops, as well as Bluetooth Low Energy interface for remote control. A mechanical integration was conducted, in the form of a controller extension which is non-intrusive and can be clipped to Valve Index controllers to enclose the TED, temperature sensors and electronics. Finally, we present a user study validating StimulHeat for use in Virtual Reality, utilizing a Unity-built virtual environment with our open-source package.
}\\



\newpage

\textbf{Specifications table}
\vskip 0.2cm
\renewcommand{\arraystretch}{1.5} 
\begin{tabularx}{\linewidth}{|>{\hsize=0.7\hsize}X|>{\hsize=1.3\hsize}X|}
  \hline \textbf{Hardware name} & {StimulHeat}
  \\
  \hline \textbf{Subject area} & {Educational tools and open-source alternatives to existing infrastructure}
  \\
  \hline \textbf{Hardware type} & {Haptic System for Thermal Perception in Virtual Reality}
  \\ 
  \hline \textbf{Closest commercial analog} & {TouchDIVER Pro gloves from WEART}
  \\
  \hline \textbf{Open source license} & {Hardware: CERN-OHL-S v2\newline Firmware and Software: GPLv3}
  \\
  \hline \textbf{Cost of hardware} & {$ \simeq 400 \$ $}
  \\
  \hline \textbf{Source file repository} & { Design files uploaded on GitHub\newline \href{https://github.com/sensors-inl/StimulHeat-Toolkit?tab=readme-ov-file\#stimulheat-project-toolkit--repository-index}{\underline{StimulHeat Project Toolkit}}}
\\\hline
\end{tabularx}

\section{Hardware in context}

\noindent In recent years, there has been considerable interest in the potential of thermal feedback for a wide range of future applications in the field of virtual reality (VR)~\cite{Lee2021,Karmakar2023,daSilveira2023,Parida2021}.
From entertainment~\cite{Plijnaer2022} to education~\cite{Raimbaud2025}, the sensation of heat or cold occupies a central position. 
Two options exist for providing thermal feedback: global and local stimulations~\cite {da_silveira_thermal_2023}.
The global approach is more straightforward, as it does not require a compromise between the device's size and performance~\cite{NytschGeusen2023}.
Global feedback is utilised for the generation of an ambient thermal perception, for instance, by means of fans and heat lamps~\cite{Dionisio1997,Han2019,Villenave2025}. 
In contrast, local feedback is used to create a sensation localized to a specific part of the body.
The principal challenge for local thermal feedback is to combine miniaturization of the device with efficient heat transfer.
In recent literature, several solutions have been proposed for local thermal feedback, including chemical stimulation~\cite{jiang2021,hamazaki2024}, pneumatic injection into gloves~\cite{Cai2020}, and thermoelectric excitation~\cite{keef2020,Maeda2019,Ragozin2020,Han2020,Philippe2025,Ranasinghe2017,Peiris2017,Kim2020,Mazursky2024,kang2024,Zhu2019}.

As raised in the literature, the most common technology is thermoelectric excitation based on Peltier elements, which can produce either heating or cooling with a small device size. 
However, existing devices using Peltier elements exhibit certain limitations: they require high power consumption due to voltage modulation~\cite{keef2020,Maeda2019,Philippe2025,Ranasinghe2017,Peiris2017,Kim2020,Mazursky2024,kang2024,Zhu2019,Xu2020}, only rely on temperature setpoints combined with Proportional-Integral-Derivative~(PID) controller to adjust the intensity of the stimulation~\cite{NytschGeusen2023,Ranasinghe2017,Peiris2017,Kim2020,Mazursky2024,kang2024,Zhu2019}, are complex to reproduce~\cite{keef2020,Maeda2019,Ragozin2020,Han2020,Kim2020,Mazursky2024,kang2024,Zhu2019} and may be very intrusive~\cite{Ragozin2020,Mazursky2024,kang2024,Wang2024}. 
In this context, we propose Stimulheat, a wireless local thermal feedback system based on Peltier elements. 
In contrast to most existing work, our device is based on heat flow control (rather than temperature). 
Heat flow has a direct and continuous relationship with the injected current, thus this approach eliminates the need for calibrating a PID controller, thereby increasing robustness. 
Moreover, the use of a Direct Current~(DC) source, rather than a pulse width modulation~(PWM) voltage source associated with a PID, reduces power consumption and increases the precision of the thermal response. 
Ultimately, the integration with Valve Index controllers ~\cite{controllersValveIndex} results in high usability and comfort. 
To enable the use of the device across a wide range of applications, the conventional temperature control method has also been incorporated in addition to a heat flow-based control approach. 
This allows users to select the most appropriate method based on their specific objectives.

\section{Hardware description}\label{Hardware description} 

\noindent StimulHeat is a palm-located heater/cooler system with a non-intrusive, low-power, dual control loop that is fully compatible with any virtual reality project utilising the Valve Index controllers. 
The selection of this particular controller was primarily determined by its forced, uninterrupted contact with the palm of the user, thereby ensuring constant thermal feedback on this contact surface. 
The thermoregulatory process is based on a system which incorporates a Peltier element, whereby the temperature of each of the two faces is continuously measured. 
Two selectable control loops have been implemented to either control heat flow or temperature. 
In both cases, the thermoelectric device~(TED) is driven by a custom-made current source and a dedicated electronic control board, which are integrated together in a 3D-printed model designed to be easily clipped onto the aforementioned controllers. 
StimulHeat's compact and wireless design allows for future integration with a diverse range of VR controllers, e.g., Meta Quest 3~\cite{metaQuest3} and Sony Playstation VR2~\cite{PSVR2}, or tangible interfaces for Augmented and Mixed Reality. 

\subsection{\underline{Custom-made current source characteristics}}

\noindent In order to drive a Peltier module, it is necessary to modulate the value of the current flowing through it. 
Two methods exist for achieving this: the direct injection of a continuous current, or the intermittent imposition of a voltage which induces an average current in turn. 
A review of recent literature revealed that the majority of existent systems which employ TED in VR for thermal feedback are controlled by a driver based on voltage modulation, comprising an H-bridge and utilising the PWM method to modulate the value of the voltage~\cite{Maeda2019,Philippe2025,Ranasinghe2017,Peiris2017,Kim2020,Mazursky2024,kang2024,Zhu2019,Xu2020}. Additionally, a limited number of studies have opted to utilise a filter in order to smooth the output of the PWM stage, thereby creating a modulable continuous voltage source~\cite{watkins2025}. 

\par
\begin{figure}[!htbp]
    \centering
    \includegraphics[width=\linewidth]{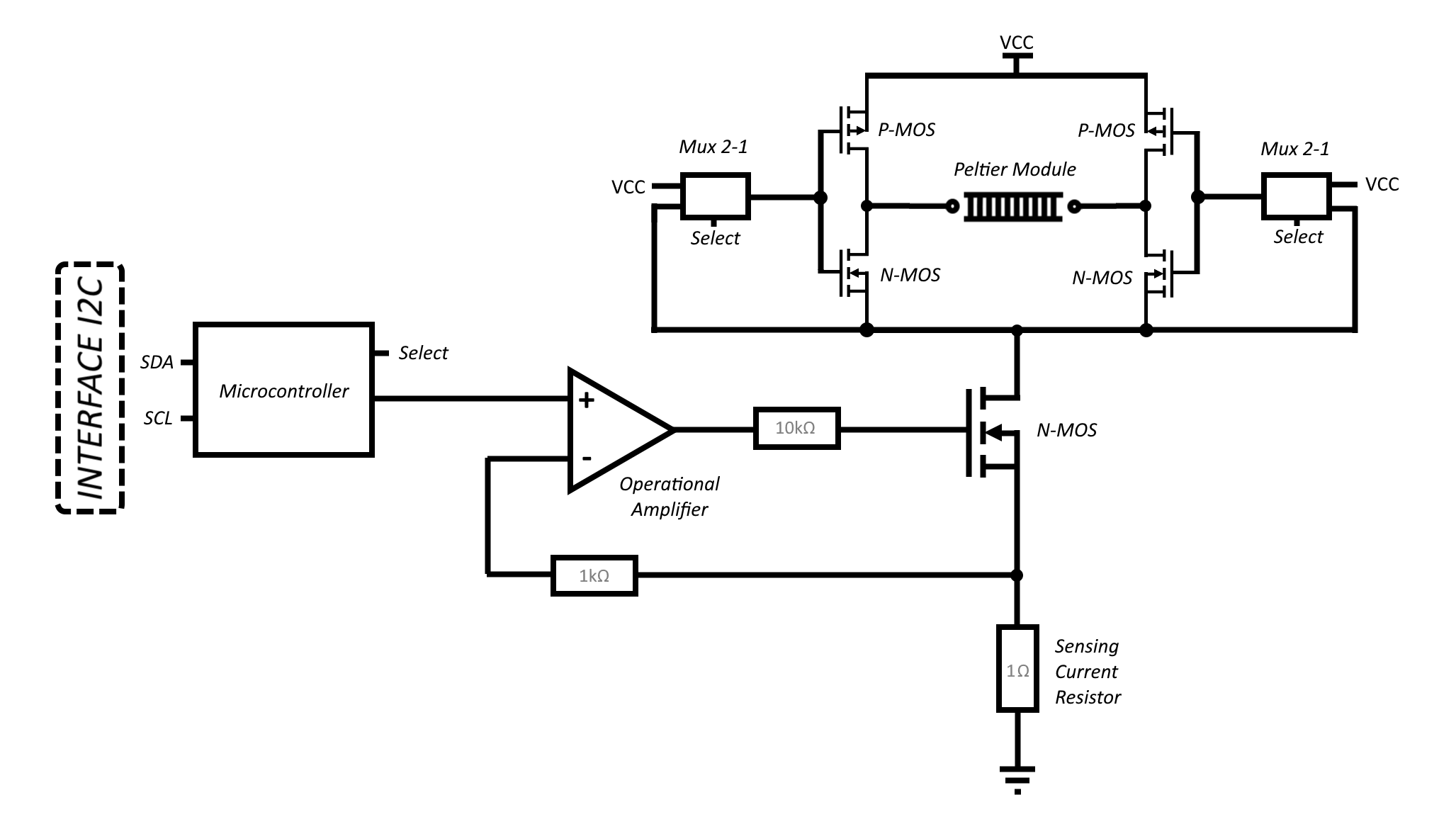}
    \caption{Electronic Design of driver based on continuous current source for a TED}
    \label{fig:flowchart_electronic_current_source}
\end{figure}

\vspace{0.5em}
\noindent A study conducted by Texas Instruments~\cite{Mellin2020} indicates that it is advisable to regulate a Peltier module with a continuous source of current rather than a PWM current generator to enhance efficiency, as a current control is better suited to the intrinsic behavior of a Peltier element. 
Consequently, we chose to drive our Peltier module with a driver entirely based on continuous current source without a PWM stage. 
The circuit diagram is shown in Figure \ref{fig:flowchart_electronic_current_source}. 
This choice of power supply benefits from the natural relationship of a Peltier element between the current injected and its ability to heat or cool through thermal conduction~\cite{goldsmid2016}.

\par\vspace{0.5em}

\noindent Firstly, a microcontroller (PIC16F17115) is responsible for the I2C communication used to receive the desired output current. 
Given the circuit's transfer function, the MCU converts this command into a DC voltage using an integrated digital-to-analog converter. 
This voltage is used at the input of an operational amplifier (LM358B) to maintain a constant voltage across a sensing current resistor (see Figure \ref{fig:flowchart_electronic_current_source}), thus defining the current that will flow through the Peltier element and the N-MOS (Si4838BDY) to which sensing current resistor is connected. 
The selection of the NMOS model is influenced by the minimum voltage gate source ($V_{GS}$) tension. 
The smaller this tension, the larger the current variation in response to the voltage delivered by the microcontroller unit (MCU). 
An H-bridge (Si7540DP and 74LVC1G3157GW125) is then fitted around the Peltier element to reverse the direction of the current flowing through it.

\subsection{\underline{Heat flow or temperature control modes}}

\par
\begin{figure}[!htbp]
    \centering
    \includegraphics[width=0.95\linewidth]{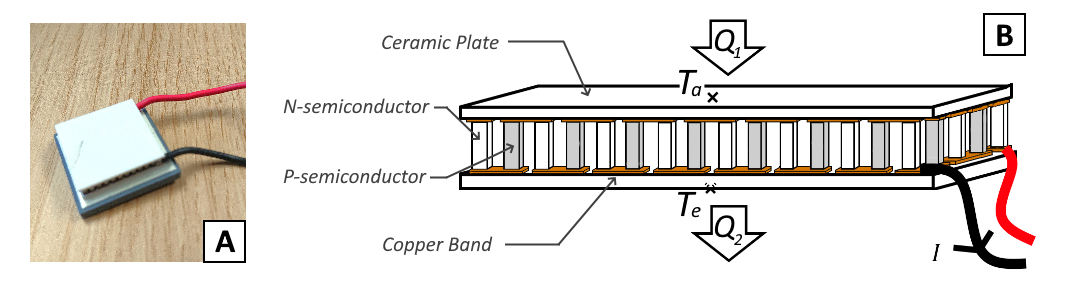}
    \caption{\textbf{A:} Picture of a TED with a ceramic heat sink - \textbf{B:} Schematic of a TED}
    \label{fig:peltier_schematic}
\end{figure}

\vspace{0.5em}
\noindent Stimulheat incorporates two methods of controlling the Peltier element, as the device is powered by a voltage-controlled, reversible, direct current source. 
Based on the theoretical equations and the operating diagram of a Peltier module shown on Figure \ref{fig:peltier_schematic}, it can be demonstrated that the heat flow passing through it is given by the relation : 

\begin{equation}
Q_1 = -\frac{R}{2} I^2 + \alpha T_a I + \frac{T_a - T_e}{\theta_m}
\label{eq:Qa}
\end{equation}

Where:
\begin{itemize}[label=--]
    \item $Q_1$ : Heat flow which occur on the absorbed side of the Peltier Module (in Watt)
    \item $R$ : Electrical resistance equivalent of the Peltier module (in Ohm)
    \item $\alpha$ : Seebeck coefficient equivalent of the Peltier module (in Volt per Kelvin)
    \item $\theta_m$ : Thermal resistivity equivalent of the Peltier module (in Kelvin per Watt)
    \item $T_a$ : Temperature of the absorbed side (in Kelvin)
    \item $T_e$ : Temperature of the emitted side (in Kelvin)
    \item $I$ : Continuous current injected (in Ampere)
\end{itemize}

\par\vspace{0.5em}
\noindent From this standpoint, two distinct approaches to control stand out. 
The initial approach involves the direct management of the value of the heat flow. 
Based on the second-order equation (\ref{eq:Qa}), it is possible to control a given heat flow by injecting a specific current value, and conversely, to determine the required current from a desired heat flow. 
By positioning the Peltier module in direct contact with the skin, the primary mechanism of heat transfer is facilitated through conduction. 
The loss of heat due to convection is minimised by ensuring effective thermal contact between the Peltier module and the skin. 
The second method involves the injection of current in accordance with the difference between a temperature setpoint and the temperature measured at the contact zone between the TED and the skin. 
The incorporation of a Proportionnal-Integral-Derivative (PID) controller serves to improve the precision and response time of the targeted temperature. 
The default values of parameters $K_p$, $K_i$ and $K_d$ of the PID controller are embedded in the firmware, yet these parameters can be modified through the wireless interface. 
The selection of each control methods depends on the specific requirements of the VR application, directly tied to the design of the interactive graspable object within the VE. 
In circumstances where the object is characterised as a continuous heat source, the utilisation of a heat flow control may be recommended. Conversely, in scenarios where the grasped object, once grasped, tends to approach a temperature close to the body's skin temperature over time, a temperature control may be more appropriate.
A technical characterisation of both control methods is presented in \hyperref[Validation and characterization]{7. Validation and Characterization} section.

\par\vspace{0.5em}
\noindent We hypothesise that heat flow control is more appropriate for thermal perception in cases where there is direct contact between an object and the skin and particularly when using a Peltier element.
First, from an energy efficiency standpoint, it is more effective to directly control heat flow rather than surface temperature $T_a$ (see Equation~\ref{eq:Qa}).
Second, when a real object is grasped and perceived, the sensation is not solely determined by temperature. Several studies have shown that even a nearly constant surface temperature can still produce distinct thermal sensations~\cite{ManasrahJun2017,ManasrahJan2017}. This phenomenon is attributed to the thermal effusivity of the object's material~\cite{Jain2023,Oss2022}.
This can be illustrated by a common observation: in a room maintained at a constant temperature of \qty{20}{\celsius}, a wooden spoon typically feels warmer than a metal one, despite both being at the same initial temperature. This occurs because steel (a metal) transfers heat more effectively to an external body than wood does—provided that the body it contacts has a lower thermal effusivity.
Consequently, the rate of temperature change during contact, until a steady-state level is attained, is higher for the steel spoon than for the wooden one. Therefore, to reproduce a realistic thermal sensation on the skin using a single material with a Peltier element, it may be more appropriate to modulate the heat flow rather than to fix the surface temperature. 
Using PID controller to maintain a specific temperature would impose a fixed rate of temperature change, which may not be suitable for replicating the sensation associated with materials of different thermal effusivities.

\par\vspace{0.5em}
\noindent To summarize, the device can be controlled in two different ways: by using a heat setpoint (in Watts) or a temperature setpoint (in Celsius). 
These utilisation of the wireless protocol implemented over Bluetooth Low Energy, using Protocol Buffers, allows for two distinct options. 
The user can select a generic setpoint from VERY HOT to VERY COLD, with five predefined levels for both heat and temperature: for heat, VERY HOT (\qty{-4}{\W}), HOT (\qty{-2}{\W}), NEUTRAL (\qty{0}{\W}), COLD (\qty{2}{\W}), and VERY COLD (\qty{4}{\W}); for temperature, VERY HOT (\SI{41}{\celsius}), HOT (\SI{38}{\celsius}), NEUTRAL (\SI{35}{\celsius}), COLD (\SI{32}{\celsius}), and VERY COLD (\SI{29}{\celsius}). 
Alternatively, a specific setpoint value can be defined, within the heat range of [\qty{-9}{W} ; \qty{9}{W}] and the temperature range of [\qty{15}{\celsius} ; \qty{42}{\celsius}]. The aforementioned ranges have been deduced from the capacity of heating and cooling of our TED (ET-071-08-15) for the heat range, and from the temperature range that does not affect the nociceptors~\cite{Hardy1952,Harrison1999} for the temperature range. In principle, both ranges are accessible; in practice, positive values of cold heat flow ($>$\qty{6}{W}) and cold temperatures ($<$\qty{25}{\celsius}) would require the addition of a high-performance heat sink system to effectively dissipate the heat produced on the opposite side of the TED when it is used for cooling (c.f. \hyperref[Limitations and Conclusion]{9. Limitations and Conclusion} section).

\subsection{\underline{Low-power solution}}

\begin{figure}[ht!]
    \centering
    \includegraphics[width=0.95\linewidth]{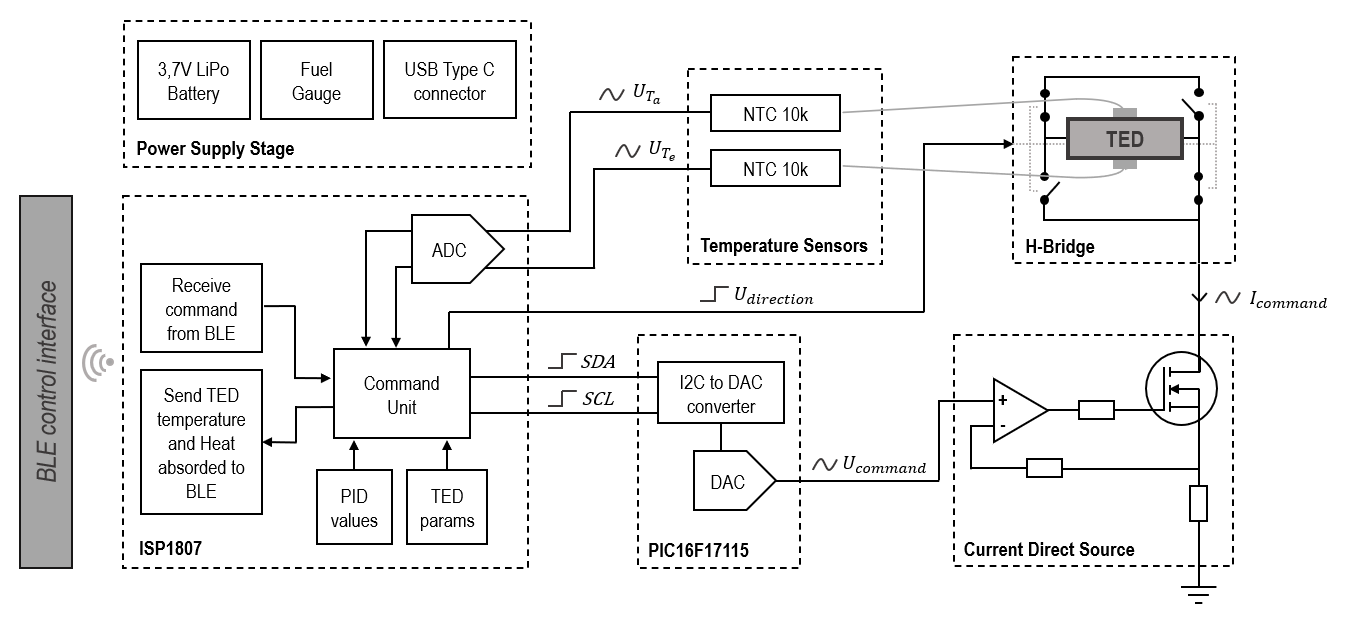}
    \caption{Flowchart of Stimulheat from the BLE command received to the current control through the Peltier module}
    \label{fig:flowchart_stimulheat}
\end{figure}

\par\vspace{0.5em}
\noindent StimulHeat is a low-power solution that employs a low-power communication protocol and a high-energy efficiency method of TED's control, making it integrable in a handheld device such as a VR controller. 
It can be used without the need for an external battery carried by the user on a distinct body site (such as the forearm, the wrist, or in a backpack), in contrast to other existing implementations~\cite{Plijnaer2022,keef2020,Kim2020,Mazursky2024}. 
The PCB and the external battery are housed within a small enclosure on a 3D printed shell clipsable on the Valve Index controller (shown in Figure \ref{fig:controller_shell_overview}). 
In order to facilitate remote control, an ISP1807 (Insight SIP), which is a nRF52804-based system-in-package (Nordic Semiconductors), was utilised for its proven low current consumption and its integrated Bluetooth Low Energy (BLE) interface, which significantly reduces energy consumption whilst ensuring interoperability and reliable command reception.

\par\vspace{0.5em}
\noindent Regarding the selection of the battery, we installed a small rechargeable LiPo battery (L.l.h : \qty{45}{mm} x \qty{34.5}{mm} x \qty{6.5}{mm}) of \qty{3.7}{V} with a capacity of \qty{850}{mAh}. 
It was determined that the application of a current of less than \qty{600}{mA} on the ET-071-08-15 series (TED from European Thermodynamics), previously selected based on three specific parameters ($\alpha$, $\theta_m$, $R$), resulted in a pronounced thermal sensation, both in the cold and in the heat (under the condition that the current is inverted in the latter case). 
In addition, the TED functions at a voltage of \qty{3.7}{V}, which results in one StimulHeat's maximum power consumption of \qty{2.22}{W}, and \qty{4.44}{W} for both controllers. 
For comparison with existing devices that provide thermal feedback to the hand or other body areas using TEDs, the Ambiotherm system from Ranasinghe \textit{et al.} consumes approximately \qty{13}{W} when using two Peltier elements for thermal stimulation at the neck~\cite{Ranasinghe2017}. The ThermalGrasp system from Mazursky \textit{et al.} operates at around \qty{25}{W} using a Peltier-based setup~\cite{Mazursky2024}. The Flip-Pelt device from Kang \textit{et al.} consumes up to \qty{9.6}{W} by using eight TEDs on the forearm~\cite{kang2024}, while the thermal display glove from Kim \textit{et al.} consumes an average of \qty{1.8}{W} using four custom-made flexible TEDs~\cite{Kim2020}.


\subsection{\underline{Integration on Valve Index controllers}}

\begin{figure}[ht!]
    \centering
    \includegraphics[width=0.5\linewidth]{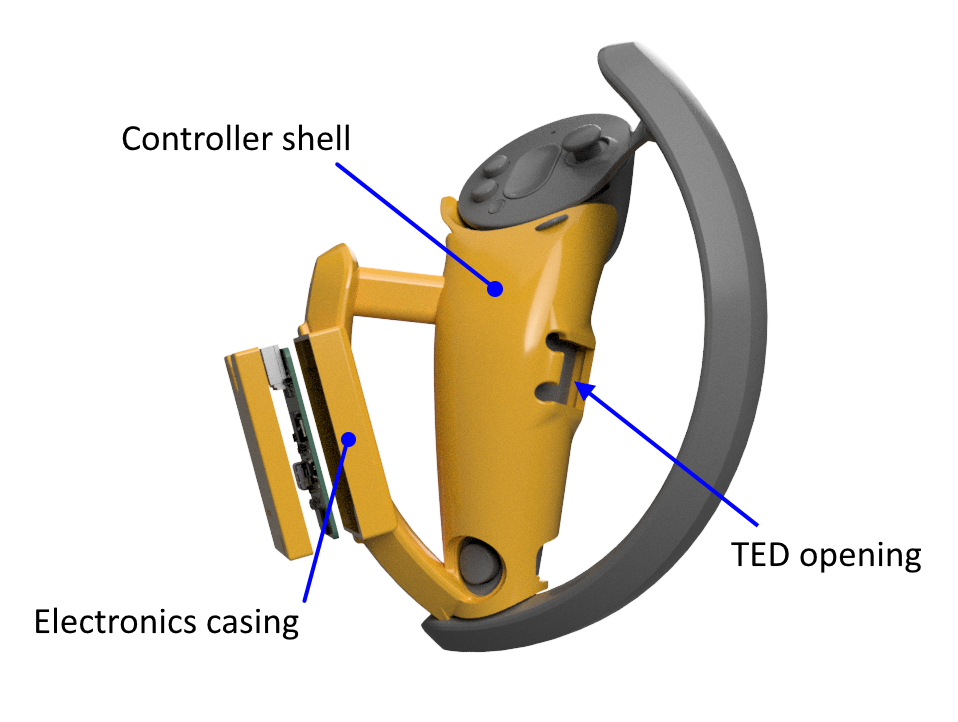}
    \caption{3D model of the accessory fitting on the Valve Index controller}
    \label{fig:controller_shell_overview}
\end{figure}

\par\vspace{0.5em}
We designed an integration module for the Valve Index controllers. Initially, the STL file provided by Valve Corporation is utilised (\href{ https://github.com/ValveSoftware/IndexHardware}{\underline{Github}}), as it is designed to be an accessory shell that can be added to a Valve Index controller. This was originally created with the intention of accommodating individuals with longer thumbs or to be used as a base for additional accessories. Figure \ref{fig:controller_shell_overview} illustrates the 3D printed custom controller shell that we created from the previous STL file. A square opening has been engineered to receive the Peltier element, situated at palm level. An extension of the controller has been incorporated to encapsulate the electronics without compromising the user's freedom of movement. Two grooves with tubes have been embedded within the 3D model to facilitate wire routing from the housing to the Peltier element.

\par\vspace{0.5em}
\noindent This integration enables any user of Valve Index controllers to utilise a haptic thermal feedback system when grasping an object in a VE. For the article purpose, the integration will be demonstrated solely on the Valve Index controllers, as the innovation lies predominantly in the StimulHeat device itself. Nevertheless, this integration can readily be adapted to other controllers, such as Meta Quest Touch~\cite{controllersmetaQuest3} or Vive Controllers~\cite{controllersvivepro2}, for example.

\par\vspace{0.5em}
\noindent In summary, by reproducing StimulHeat you will have access to:
\begin{itemize}[label=$\bullet$]
    \item An easy-to-use implementation of thermal feedback in VR applications integrated in Valve Index controllers
    \item A reactive feedback solution enabling realistic thermal sensations when grasping objects in VR, as further shown in the \hyperref[Validation and characterization]{section 7}.
    \item A non-intrusive disposition giving the feeling of freedom for the user
    \item A reproducible solution which is low-power for environmentally friendly and long-term use
    \item A possibility to adapt the system to any other controllers by only changing the 3D model of the shell
\end{itemize}

Furthermore, StimulHeat was showcased in the research demonstration track at the IEEE VR 2025 conference~\cite{Mesnage2025}, revealing significant interest from the VR community in such a device.

\section{Design files summary}
\vskip 0.1cm
\renewcommand{\arraystretch}{1.5} 
\begin{tabularx}{\linewidth}{|>{\hsize=0.75\hsize}X|>{\hsize=1.05\hsize}X|>{\hsize=0.9\hsize}X|>{\hsize=1.3\hsize}X|} 

\hline
\textbf{Design filename} & \textbf{File type} & \textbf{Open source license} & \textbf{Location of the file} \\\hline
{\scriptsize{StimulHeat-3D-Parts}} & {ZIP File (3D STEP files)} & {CERN-OHL-S v2} & {\href{https://github.com/sensors-inl/StimulHeat-3D-Parts}{\underline{Stimulheat - 3D Parts}}} \\\hline
{\scriptsize{StimulHeat-Hardware}} & {ZIP File (ECAD files)} & {CERN-OHL-S v2} & {\href{https://github.com/sensors-inl/StimulHeat-Hardware}{\underline{Stimulheat - Hardware}}} \\\hline
{\scriptsize{StimulHeat-Firmware-PIC16F17115}} & {ZIP File (C source code)} & {GPLv3} & {\href{https://github.com/sensors-inl/StimulHeat-Firmware-PIC16F17115}{\underline{Stimulheat - Firmware PIC16F}}} \\\hline
{\scriptsize{Stimulheat-Firmware-ISP1807}} & {ZIP File (C source code)} & {GPLv3} & {\href{https://github.com/sensors-inl/StimulHeat-Firmware-ISP1807}{\underline{Stimulheat - Firmware ISP1807}}} \\\hline
{\scriptsize{StimulHeat-WebApp}} & {ZIP File (HTML/JS files)} & {GPLv3} & {\href{https://github.com/sensors-inl/StimulHeat-WebApp}{\underline{Stimulheat - Web App}}} \\\hline
{\scriptsize{StimulHeat-Unity}} & {ZIP File (Unity project)} & {GPLv3} & {\href{https://github.com/sensors-inl/StimulHeat-Unity}{\underline{Stimulheat - Unity Package}}} \\\hline
\end{tabularx}


\begin{itemize}
\item{StimulHeat-3D-Parts: ZIP File which contains the printable 3D models in STEP and STL formats (for SLA or FDM).}
\item{StimulHeat-Hardware: ZIP File which contains the entire KiCAD project, including the schematic (stimulheat.kicad\_sch file), the PCB (stimulheat.kicad\_pcb file), the symbol library (stimulheat.kicad\_sym file), the footprint library (stimulheat.pretty folder) and the 3D model of the components library (stimulheat-3dmodels folder). Also manufacturing (GERBER and DRILL) files, assembly data (component position placing file, assembly drawing, bill of materials) are provided for a complete manufacturing and assembly of the printed circuit board.}
\item{StimulHeat-Firmware-PIC16F17115: ZIP File which contains all the source code to program the PIC16F17115 on the StimulHeat PCB. A prebuilt binary file is also provided for direct programming of the microcontroller.}
\item{StimulHeat-Firmware-ISP1807: ZIP File which contains all the source code needed to program the ISP1807 on the StimulHeat PCB. A prebuilt binary file is also provided for direct programming of the microcontroller.}
\item{Stimulheat-WebApp: ZIP file which contains the local standalone HTML/JS web application allowing you to test your StimulHeat device.}
\item{StimulHeat-Unity: ZIP file which contains the Unity virtual demo scene used in the \hyperref[Validation and characterization]{7. Validation and Characterization} section.}
\end{itemize}

\section{Bill of materials summary}


\renewcommand{\arraystretch}{1.5}
\begin{tabularx}{\linewidth}{|>{\hsize=1.1\hsize}X|>{\hsize=1.0\hsize}X|>{\hsize=0.6\hsize}X|>{\hsize=0.8\hsize}X|>{\hsize=1.0\hsize}X|>{\hsize=1.0\hsize}X|>{\hsize=1.5\hsize}X|}

\hline
\textbf{Designator} & \textbf{Component} & \textbf{Qty} & \textbf{Cost per unit - currency} & \textbf{Total cost - currency} & \textbf{Source of materials} & \textbf{Material type} \\\hline

\scriptsize{Assembled StimulHeat PCB}\label{BOM} & 
\scriptsize{StimulHeat} & 
\scriptsize{1} & 
\scriptsize{\(\sim 150~ \$\)} & 
\scriptsize{\(\sim 150~ \$\)} & 
\scriptsize{DIY or  \href{https://www.eurocircuits.com/}{\underline{Eurocircuit}}} &
\scriptsize{Others} \\\hline

\scriptsize{3D-printed - Main} & 
\scriptsize{3D Print-Main} & 
\scriptsize{1} & 
\scriptsize{\(\sim 20~ \$\)} & 
\scriptsize{\(\sim 20~ \$\)} &
\scriptsize{DIY or \href{https://jlcpcb.com/}{\underline{JLCPCB}}} & 
\scriptsize{Others} \\\hline

\scriptsize{3D-printed - PCB Housing} & 
\scriptsize{3D Print - PCB} & 
\scriptsize{1} & 
\scriptsize{\(\sim 5~ \$\)} & 
\scriptsize{\(\sim 5~ \$\)} & 
\scriptsize{DIY or \href{https://jlcpcb.com/}{\underline{JLCPCB}}} & 
\scriptsize{Others} \\\hline

\scriptsize{3D-printed - TED Support} & 
\scriptsize{3D Print - TED Support} & 
\scriptsize{1} & 
\scriptsize{\(\sim 2~ \$\)} & 
\scriptsize{\(\sim 2~ \$\)} & 
\scriptsize{DIY or \href{https://jlcpcb.com/}{\underline{JLCPCB}}} & 
\scriptsize{Others} \\\hline

\scriptsize{PicKit5} & 
\scriptsize{PG164150} & 
\scriptsize{1} & 
\scriptsize{74.90 \$} & 
\scriptsize{74.90 \$} & 
\scriptsize{ \href{https://octopart.com/fr/pg164150-microchip-136159506}{\underline{Octopart}}} & 
\scriptsize{Others} \\\hline

\scriptsize{Tag-Connect 6-Pin - PicKit5} & 
\scriptsize{TC2030-PKT-NL} & 
\scriptsize{1} & 
\scriptsize{\(43.00~ \$\)} & 
\scriptsize{\(43.00~ \$\)} & 
\scriptsize{ \href{https://www.tag-connect.com/product/tc2030-pkt-nl-6-pin-no-legs-cable-for-microchip-pickit-3}{\underline{Tag-Connect}}} & 
\scriptsize{Others} \\\hline

{\raggedright \scriptsize{J-LINK EDU MINI}\par} & 
\scriptsize{8.08.91} & 
\scriptsize{1} & 
\scriptsize{60.00 \$} & 
\scriptsize{60.00 \$} & 
\scriptsize{ \href{https://octopart.com/fr/8.08.91-segger-84258703}{\underline{Octopart}}} & 
\scriptsize{Others} \\\hline

\scriptsize{Tag-Connect 6-Pin - ISP1807} & 
\scriptsize{TC2030-CTX-NL} & 
\scriptsize{1} & 
\scriptsize{\(42.95~ \$\)} & 
\scriptsize{\(42.95~ \$\)} & 
\scriptsize{ \href{https://www.tag-connect.com/product/tc2030-ctx-nl-6-pin-no-legs-cable-with-10-pin-micro-connector-for-cortex-processors}{\underline{Tag-Connect}}} & 
\scriptsize{Others} \\\hline

\scriptsize{Clip to hold Tag-Connect 6-Pin} & 
\scriptsize{TC2030-CLIP-3PACK} & 
\scriptsize{1} & 
\scriptsize{\(18.00~ \$\)} & 
\scriptsize{\(18.00~ \$\)} & 
\scriptsize{ \href{https://www.tag-connect.com/product/tc2030-retaining-clip-board-3-pack}{\underline{Tag-Connect}}} & 
\scriptsize{Others} \\\hline

\scriptsize{3.7V-850mAh LiPo Battery} & 
\scriptsize{LP603443JU} & 
\scriptsize{1} & 
\scriptsize{9.17 \$} & 
\scriptsize{9.17 \$} & 
\scriptsize{\href{https://octopart.com/fr/search?q=LP603443JU&currency=USD&specs=0}{\underline{Octopart}}} & 
\scriptsize{Others} \\\hline

\scriptsize{Optopipes} & 
\scriptsize{51513020250F} & 
\scriptsize{3} & 
\scriptsize{1.01 \$} & 
\scriptsize{3.03 \$} & 
\scriptsize{\href{https://octopart.com/fr/515-1302-02-50f-dialight-50829707}{\underline{Octopart}}} & 
\scriptsize{Others} \\\hline

\scriptsize{Thermistors} & 
\scriptsize{GA10K3MCD1} & 
\scriptsize{2} & 
\scriptsize{13.73 \$} & 
\scriptsize{27.46 \$} & 
\scriptsize{\href{https://octopart.com/fr/ga10k3mcd1-te+connectivity-59208877}{\underline{Octopart}}} & 
\scriptsize{Others} \\\hline

\scriptsize{TED} & 
\scriptsize{ET-071-08-15} & 
\scriptsize{1} & 
\scriptsize{\( 46.30~ \$\)} & 
\scriptsize{\( 46.30~ \$\)} & 
\scriptsize{\href{https://octopart.com/fr/et-071-08-15-rs-european+thermodynamics-119059521}{\underline{Octopart}}} & 
\scriptsize{Others} \\\hline

\scriptsize{Heat Sink} & 
\scriptsize{MPC202025T} & 
\scriptsize{1} & 
\scriptsize{1.10 \$} & 
\scriptsize{1.10 \$} & 
\scriptsize{\href{https://octopart.com/fr/mpc202025t-amec+thermasol-24435314}{\underline{Octopart}}} & 
\scriptsize{Others} \\\hline

\scriptsize{M2 Screws} & 
\scriptsize{RM2X8MM 2701} & 
\scriptsize{2} & 
\scriptsize{0.85 \$} & 
\scriptsize{1.70 \$} & 
\scriptsize{\href{https://octopart.com/fr/rm2x12mm+2701-apm+hexseal-24902804}{\underline{Octopart}}} & 
\scriptsize{Others} \\\hline

\scriptsize{M2 Nuts} & 
\scriptsize{MHNZ 002 4} & 
\scriptsize{2} & 
\scriptsize{0.10 \$} & 
\scriptsize{0.20 \$} & 
\scriptsize{\href{https://octopart.com/fr/mhnz+002+4-building+fasteners-22099385}{\underline{Octopart}}} & 
\scriptsize{Others} \\\hline

\scriptsize{Wiring Harness} & 
\scriptsize{2162711060} & 
\scriptsize{1} & 
\scriptsize{2.77 \$} & 
\scriptsize{2.77 \$} & 
\scriptsize{\href{https://octopart.com/fr/216271-1060-molex-112847899}{\underline{Octopart}}} & 
\scriptsize{Others} \\\hline

\scriptsize{Heat-shrinkable Tube \qty{1}{\milli\meter}} & 
\scriptsize{RNF-3000-3/1-0-SP} & 
\scriptsize{$\simeq$ 1 meter} & 
\scriptsize{3.12 \$} & 
\scriptsize{3.12 \$} & 
\scriptsize{\href{https://octopart.com/fr/rnf-3000-3\%2F1-0-sp-te+connectivity+\%2F+raychem-41928991}{\underline{Octopart}}} & 
\scriptsize{Others} \\\hline

\scriptsize{Heat-shrinkable Tube \qty{0,5}{\milli\meter}} & 
\scriptsize{RNF-3000-1.5/0.5-0-SP} & 
\scriptsize{$\simeq$ 1 meter} & 
\scriptsize{1.15 \$} & 
\scriptsize{1.15 \$} & 
\scriptsize{\href{https://octopart.com/fr/search?q=RNF-3000-1.5\%2F0.5-0-SP&currency=USD&specs=0}{\underline{Octopart}}} & 
\scriptsize{Others} \\\hline

\end{tabularx}\\

\noindent \underline{Additional descriptions - Global:} Please find below a list of materials that may not already be included in the previous list. 
These are workbench / assembly tools which can be useful to help assembling the device manually as it is described in the Build Instructions section :
\begin{itemize}
\item{A wire stripper to remove the insulation from electrical wires - \href{https://octopart.com/fr/63817-0000-molex-828306}{\underline{Octopart}} - Price : 120,19 \$}
\item{A hot air iron to heat the heat shrink tubing, enabling it to be retracted on itself. This, in turn, ensures electrical isolation and improves mechanical resistance. - \href{https://octopart.com/fr/ao852a\%2B\%2B220v-sra+soldering+products-116506736}{\underline{Octopart}} - Price : 159.14 \$}
\item{A soldering iron to solder component together - \href{https://octopart.com/fr/wlirk6012a-weller-118956393}{\underline{Octopart}} - Price : 51.25 \$}
\item{A polyimide adhesive tape (or any other tape that can resist to heat) to fix the thermistors on TED sides - \href{https://octopart.com/fr/kapton-tape10mm-olimex-129955975}{\underline{Octopart}} - Price : 2.84 \$}
\end{itemize}

\noindent\underline{Additional descriptions - StimulHeat PCB:} There are two options available to obtain an assembled PCB for the electronic board. 
You can either manufacture the PCB yourself or outsource its production. 
We are recommending the second option, given the complexity of some component packages, if you only have a soldering iron. 
Here is a list of subcontractors able to reproduce the project : \href{https://jlcpcb.com/}{\underline{JLCPCB}}, \href{https://www.pcbway.com//}{\underline{PCBWay}}  (Worldwide), \href{https://www.eurocircuits.com/}{\underline{EuroCircuits}}, \href{https://www.proto-electronics.com/}{\underline{Proto-Electronics}} (European) and \href{https://emsfactory.com/}{\underline{EMS FACTORY}}  (France). In order to proceed, please provide the folder 'stimulheat\_export' which can be found within the 'stimulheat\_pcb' ZIP file.

\noindent \underline{Additional descriptions - 3D Print-(all):} As with the StimulHeat PCB, you can print the 3D model yourself or use an external subcontractor. 
For the realisation of this project, we worked with \href{https://jlcpcb.com/}{\underline{JLCPCB}} to print the differents 3D model using SLA technology. 
Please be advised that there are other subcontractors, as for instance \href{https://www.hubs.com/ }{\underline{Protolabs Network}}, \href{https://www.shapeways.com/ }{\underline{Shapeways}}, \href{https://www.protolabs.com/}{\underline{Protolabs}}, etc...

\noindent{\underline{Additional descriptions - TED:} The TED selected in the materials bill is the result of a Peltier benchmark established by our own research. 
It is important to note that the utilisation of an alternative product from an alternative supplier is permissible and should not result in a significant change to the performance of StimulHeat. 
Adapting your Web or VR applications for use with your selected TED only requires changing the $\alpha_m$, $\theta_m$ and $R_m$ parameters accordingly to your TED specifications.
However, it is recommended that a similar product is employed.
The following is a concise, non-exhaustive list of Peltier elements that may be utilised as an alternative to the one that has been proposed: CP20251 - (\href{https://octopart.com/fr/search?q=CP20251&currency=USD&specs=0}{\underline{Octopart}}) ; CP30239H (\href{https://octopart.com/fr/search?q=CP30239H&currency=USD&specs=0}{\underline{Octopart}}) ; CP20247H  (\href{https://octopart.com/fr/search?q=CP20247H&currency=USD&specs=0}{\underline{Octopart}}) ; ...

\section{Build instructions}

\noindent To build the project as described below, components listed in the bill of materials are required. 
The process consists of four main steps which have to be repeated for both left and right controllers: 1) assembling the wiring harness, which includes the TED and two thermistors; 2) installing the LiPo rechargeable battery; 3) programming the microcontrollers; and finally, 4) completing the full assembly of the system.

\subsection{\underline{Creation of the Wiring Harness Containing the TED and the two Thermistors}}

To complete this step, you will need the \hyperref[BOM]{3D print-Main}, the \hyperref[BOM]{thermistors}, the \hyperref[BOM]{TED}, the \hyperref[BOM]{heat dissipator}, the \hyperref[BOM]{wiring harness} and the \hyperref[BOM]{heat shrink tubing} as it is shown in Figure \ref{fig:building_instructions_wiring_harness} - A.

\begin{itemize}

\item{First, mount the heat sink onto the TED, ensuring it is properly centered. Next, based on the length of the TED’s electrical wires, trim the two central wires of the wiring harness so that the combined length of the TED wires and the harness wires is exactly \qty{13.5}{cm}. 
Cut the thermistors to a length of \qty{7.5}{cm}, and trim the last four wires of the wiring harness to leave \qty{1}{cm} exposed on each. 
Finally, strip the insulation from the TED wires using a wire stripper. 
The final setup should match the configuration shown in Figure \ref{fig:building_instructions_wiring_harness} - B.}

\item{Then, cut six pieces of \qty{1}{mm} - diameter heat shrink tubes, each \qty{1.5}{cm} long. 
Slide two of them onto the two central wires of the wiring harness and reserve the remaining four for the thermistor wires. 
Next, pass the TED wires through the two holes in the 3D print – Main, as shown in Figure  \ref{fig:building_instructions_wiring_harness} - C. 
Solder the TED wires to the two central wires of the wiring harness. 
After soldering, cover the joints with the heat shrink tubies and shrink them using a hot air gun to ensure proper electrical insulation. 
The resulting configuration should look like the Figure \ref{fig:building_instructions_wiring_harness} - D}.

\item{Prepare the two thermistors as illustrated in Figure \ref{fig:building_instructions_wiring_harness} - E, by gently separating their two wires over a length of approximately \qty{1.5}{cm}. 
Although the insulation is already removed, handle the wires carefully as they are very thin and fragile. To reinforce them, cut two pieces of 0.6 cm heat shrink tubing, each \qty{7}{cm} in length. 
Slide them on the thermistor and shrink them using a hot air gun. 
Strip the last four wires of the wiring harness, then carefully solder the thermistor wires to their corresponding wires on the harness. 
On one of the two solder joints per thermistor, place a piece of Kapton tape to provide additional electrical insulation. 
Then, slide the remaining four \qty{1}{mm} thermal sheaths over the solder joints and shrink them using a hot air gun to ensure proper insulation. 
Verify that the two solder joints on each thermistor do not touch each other by measuring the resistance between the two thermistor wires. 
A reading of 0 ohm indicates a short circuit, in which case the step must be redone.}

\item{Once the previous step is complete, route the wires through the tubes of the 3D print – Main, following the configuration shown in Figure \ref{fig:building_instructions_wiring_harness} -  F. 
Finally, tape the thermistor onto the face of the TED using Kapton tape. Ensure that the thermistor connected to the left side of the wiring harness is attached to the face of the TED with the heat sink.}

\end{itemize}

\begin{figure}[ht!]
    \centering
    \includegraphics[width=\linewidth]{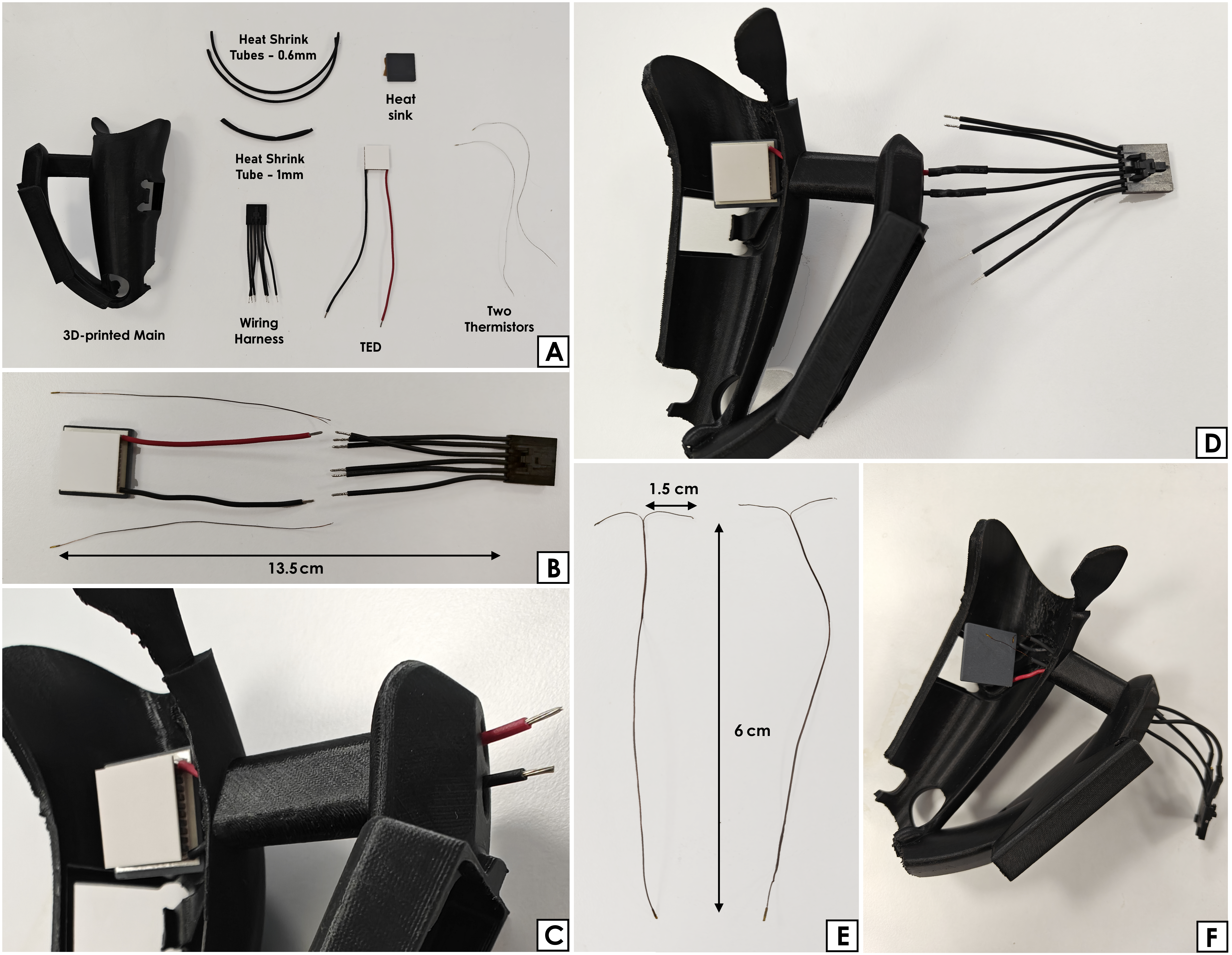}
    \caption{Building instructions - Wiring Harness}
    \label{fig:building_instructions_wiring_harness}
\end{figure}

\subsection{\underline{Installation of the Rechargeable Battery}}

For this step, you will need the  \hyperref[BOM]{StimulHeat PCB} and the \hyperref[BOM]{\qty{3.7}{V}-\qty{850}{mAh} LiPo Battery}.

\begin{itemize}

\item{Using a soldering iron, solder the black wire to the GND pad on the bottom side of the PCB, which is one of the three pads of the battery connector. 
Next, solder the red wire to the pad marked with an arrow. 
Be sure to follow this connection sequence to avoid any electrical risks. 
At the end of this step, your assembly should look similar to the example shown in Figure \ref{fig:building_instructions_installation_rechargeable_battery} - A.}

\item{To check if the soldering is done properly, try charging your battery with a USB-C cable. 
If a yellow LED light comes on, as shown in Figure \ref{fig:building_instructions_installation_rechargeable_battery} - B, this means that the device is well connected to the battery and working properly. 
If not, check your welding. The yellow LED will turn green when the battery is fully charged.}

\end{itemize}

\begin{figure}[ht!]
    \centering
    \includegraphics[width=0.75\linewidth]{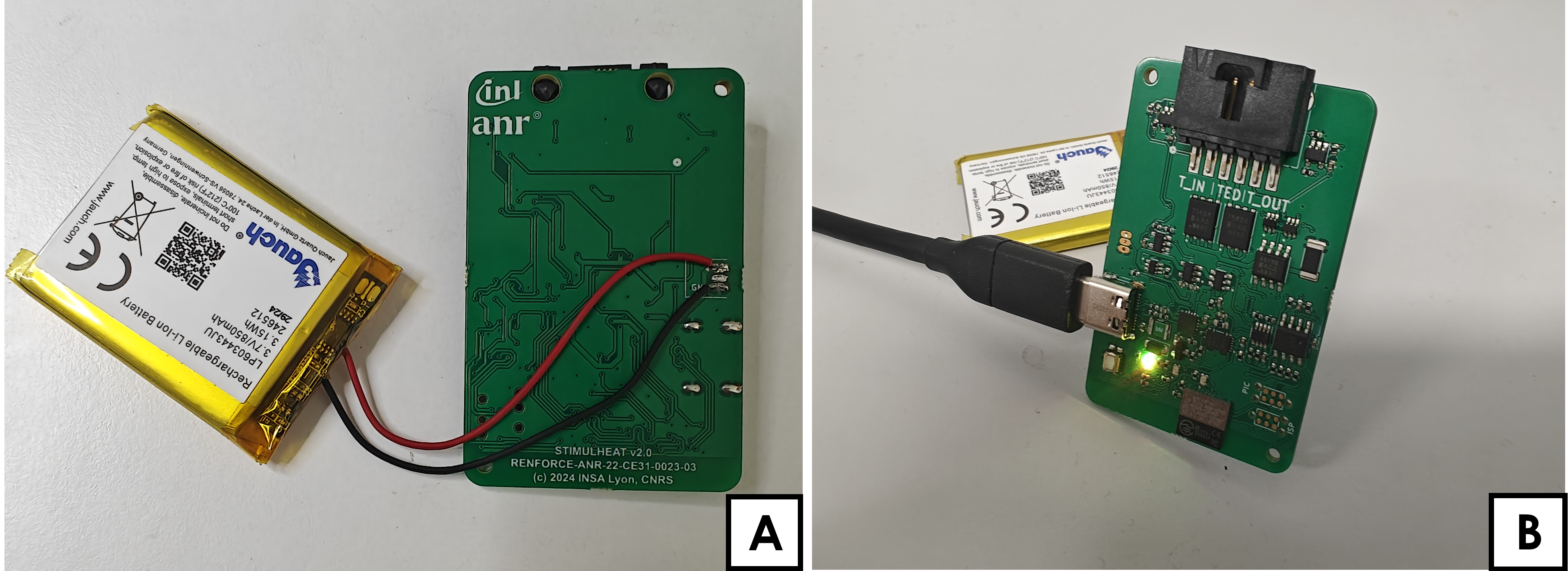}
    \caption{Building instructions - Rechargeabable Battery Installation}
    \label{fig:building_instructions_installation_rechargeable_battery}
\end{figure}

\subsection{\underline{Programming of the micro-controllers}}

The next step consists in programming the two microcontrollers on the PCB. 
You will need the previous assembly of the StimulHeat and the rechargeable battery, as well as the \hyperref[BOM]{Pickit5}, the \hyperref[BOM]{Tag-Connect 6-Pin - PicKit5}, the \hyperref[BOM]{J-LINK EDU MINI} and the \hyperref[BOM]{Tag-Connect 6-Pin - ISP1807}.

\begin{itemize}

\item{Start by charging your battery. 
The charging LED will turn yellow if the battery is not fully charged. 
Once the charging LED indicator turns green, programming can be done. 
Turn on your StimulHeat using the push button.}

\item{We begin with the PIC16F by using the Pickit5 and the Tag-Connect 6-Pin - PicKit5 programming cable. 
To do so, first connect the adapter to your PicKit5, ensuring the MCLR pin on the PicKit5 (indicated by an arrow) matches the blue marked wire on the Tag-connect cable. 
Next, connect the other end of the programming cable to the 'PIC16F' connector on the PCB. 
Finally, plug the USB cable into the PicKit5 and your computer, then open MPLAB X IPE. 
Load the hex file '\textbf{stimulheat-firmware-pic16f17115.production.hex}' (located in stimulHeat\_pic16f\_code.zip).
Select All Families for Family, PIC16F17115 for device and PicKit5 for your tool. 
Then click on Program. 
At this time, the PIC16F17115 is programmed.}

\item{Now, let's proceed with programming the ISP1807. For this step, you will need the J-LINK EDU MINI with its programming cable, the Tag-Connect 6-Pin - ISP1807. 
Start by connecting the cable to the J-LINK EDU MINI, then connect the other end to the 'ISP1807' connector on the PCB.
Finally, connect the USB cable to the J-LINK EDU MINI and your computer, then open nRF Connect for Desktop. 
Launch it, find the "Programmer" App, install it and then open it. 
On it, click on 'select device' and select J-LINK. 
By clicking on add file in File menu, select the hex file '\textbf{zephyr.hex}' (located in the ‘build/zephyr’ directory of the archive  stimulHeat\_isp1807\_code.zip). 
Then, click on 'erase and write' in Device menu. 
At this time, the ISP1807 is programmed.}

\end{itemize}

\subsection{\underline{Assembly of the whole project}}

To assemble the entire project, you will need the first sub-assembly containing the wiring harness, and the second sub-assembly composed of the StimulHeat module and the battery. 
You will also need the \hyperref[BOM]{3D-printed PCB housing}, the \hyperref[BOM]{optopipes}, and the \hyperref[BOM]{M2 screws}.

\begin{itemize}

\item{On the 3D-printed PCB housing, gently insert the three optopipes into the corresponding holes, as shown in Figure \ref{fig:building_instructions_assembly_of_the_whole_project} - A.}

\item{Then, insert the StimulHeat PCB + battery assembly into the dedicated slot on the main 3D-printed structure from the first sub-assembly. 
To help with orientation, make sure the connector for the wiring harness on the StimulHeat is positioned at the top similar to the example shown in Figure \ref{fig:building_instructions_assembly_of_the_whole_project} - B.}

\item{Insert the two M2 nuts into the designated holes on the Wiring Harness assembly. 
Then, take the 3D Print–PCB + optopipe assembly, align it with the previous structure, and secure it in place using the two M2 screws at both ends of the PCB. 
If necessary, hold the M2 nuts in position from the top side while tightening the screws to ensure proper engagement, as shown in Figure \ref{fig:building_instructions_assembly_of_the_whole_project} - C. To confirm correct installation, press the button on the 3D Print–PCB component. 
If one of the three LEDs turns blue and another one starts blinking like shown in the Figure \ref{fig:building_instructions_assembly_of_the_whole_project} - D, the assembly has been successfully installed.}

\item{If the test is successful, remember to press the button again to turn off the StimulHeat module.}

\item{Finally, position the TED module and the wires respectively into the hole and the designated cable channels on the main 3D-printed part.}

\item{Before clipping the assembly onto a Valve Index controller, don’t forget to place the 3D Print-TED part over the TED + heat sink to ensure optimal contact between the hand and the TED.}

\end{itemize}

\begin{figure}[ht!]
    \centering
    \includegraphics[width=0.5\linewidth]{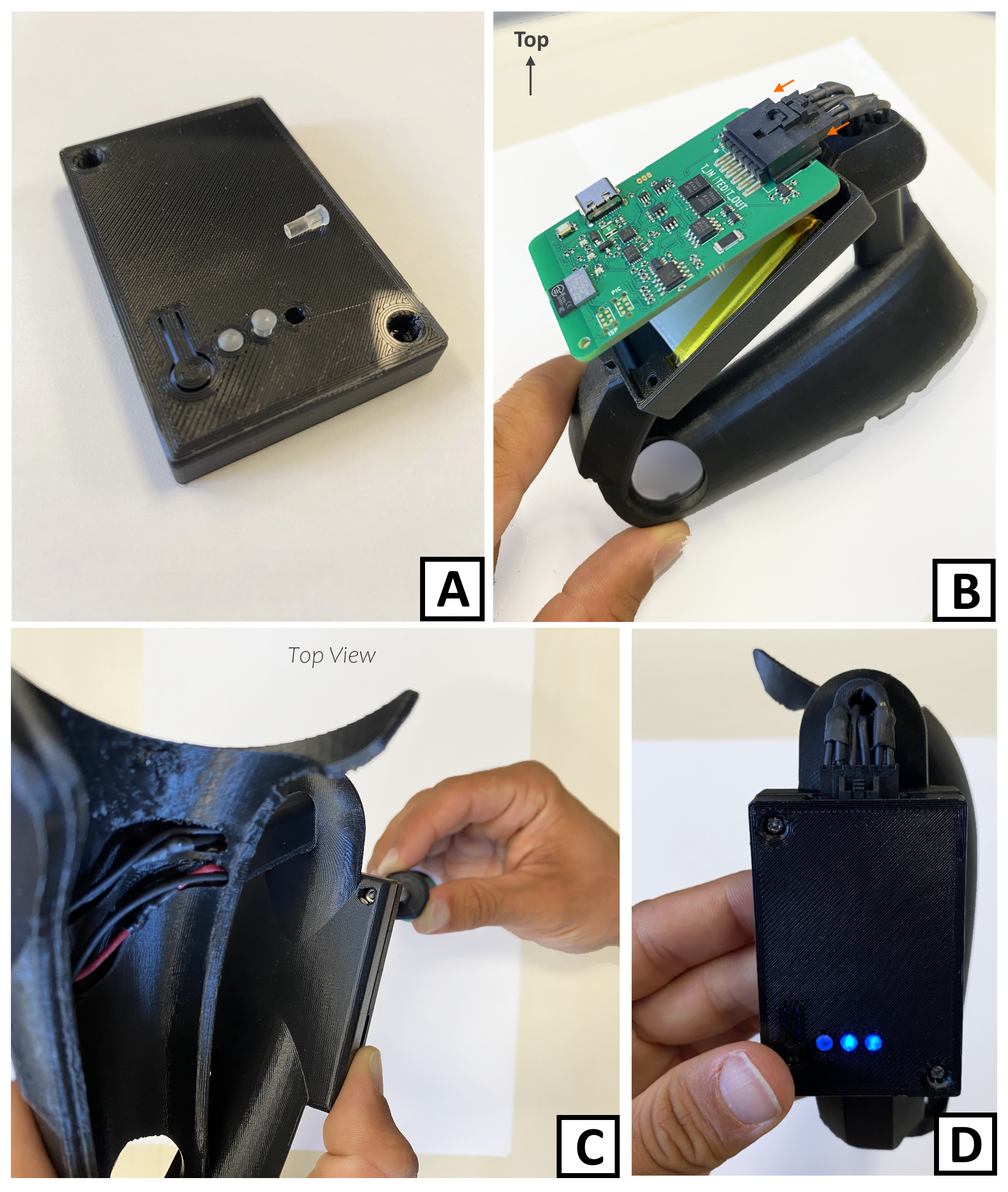}
    \caption{Building instructions - Assembly of the whole project}
    \label{fig:building_instructions_assembly_of_the_whole_project}
\end{figure}

\section{Operation instructions}

To ensure the Stimulheat operates correctly, we have organised the instructions into two parts: Device Install/Uninstall on Valve Index controller and Device Connection Protocol.

\subsection{\underline{StimulHeat Install/Uninstall on Valve Index controller}}

To install or uninstall StimulHeat on the Valve Index controller, take care to move the bottom part in order to not break the 3D print. In detailled, you can follow the sequence presented in Figure \ref{fig:operation_instructions_install-uninstall_stimulheat}.

\begin{figure}[ht!]
    \centering
    \includegraphics[width=\linewidth]{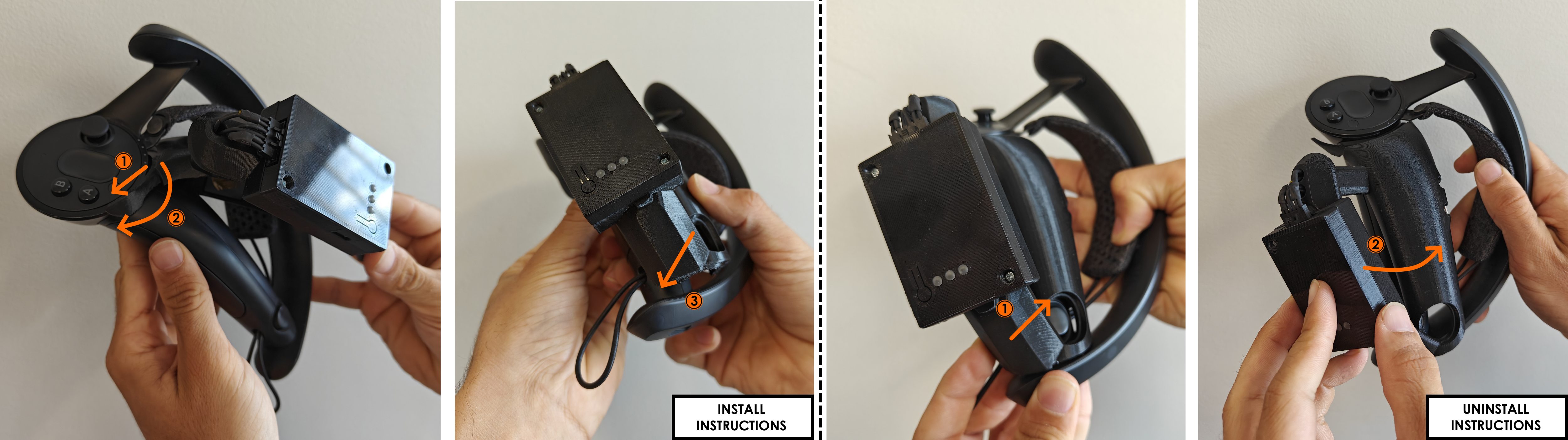}
    \caption{Operation instructions - Install-Uninstall StimulHeat on a Valve Index controller}
    \label{fig:operation_instructions_install-uninstall_stimulheat}
\end{figure}

\subsection{\underline{StimulHeat Connection Protocol - Web App}}

StimulHeat communicates with its environment via the ISP1807 module, which integrates a Bluetooth Low Energy (BLE) chip.
It uses the Google Protocol Buffers (ProtoBuf) protocol to encode and decode messages. 
The corresponding .proto file is included in the stimulheat\_isp1807\_code.zip archive and can be used to integrate StimulHeat with any application that supports Bluetooth communication. 
To help you test your hardware, we have developed a local web application that allows you to control your StimulHeat device. 
Below, you’ll find a step-by-step tutorial on how to use the local web app. 

\begin{itemize}

\item{First, download the 'stimulheat\_local\_web\_application.zip' and extract all the files. 
Then, open with a web browser (not compatible with Firefox) the 'index.html' file. 
You should find a web page similar to the one shown in Figure \ref{fig:operation_instructions_connection_protocol}}

\item{Put on your StimulHeat and click on 'CONNECT BLE'. 
You may find it in the list in the format 'StimulHeat XXXX' with XXXX the serial number of your device. 
After selecting it, click on 'Connect'.}

\item{You are now connected. 
To acess the command menu, you must enable the thermal control of the device by clicking on 'ON'. 
After that, you can change the value of the command as you want. 
To stop the use of the StimulHeat, click on 'OFF'}

\end{itemize}

\begin{figure}[ht!]
    \centering
    \includegraphics[width=1.0\textwidth]{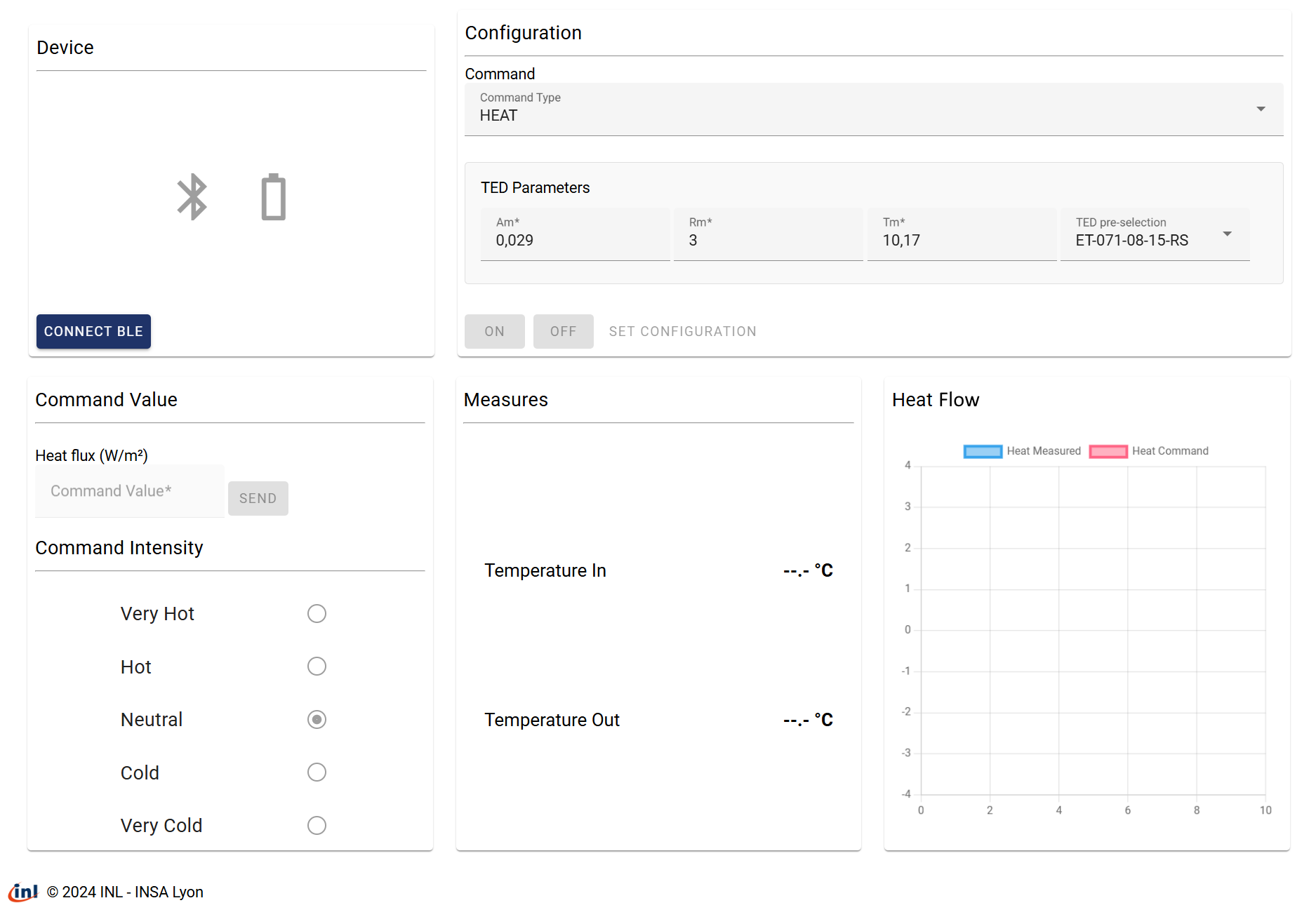}
    \caption{Operation instructions - Web App (screenshot)}
    \label{fig:operation_instructions_connection_protocol}
\end{figure}

\textbf{Important note :} At this stage, you may notice that the command response is inverted relative to the intended behavior. 
Do not worry—this simply indicates that the TED has been connected in reverse. 
To correct this, cut, strip, and swap the two relevant wires of the wiring harness before reconnecting them to the TED. 
This inversion will allow you to control the StimulHeat correctly via the web application.

\subsection{\underline{StimulHeat Connection Protocol - Unity}}

To enable seamless integration of the Stimulheat device with Unity\footnote{\url{https://unity.com/}} based applications, we developed a dedicated package easily installable via Unity’s Package Manager, using the provided Git repository link. This package also bundles a desktop demo scene in the samples, they can be added directly in your Unity project from the Package Manager. After installation within your Unity Project, you can quickly configure and control the device by starting the bundled demo scene, located in \textit{Assets/Samples/StimulHeat/0.0.1/Demo} (provided your computer is equipped with Bluetooth capabilities). If you don't have Unity installed on your computer, you can also try out the demo application by launching the provided build. Once launched, the configuration interface (see Figure~\ref{fig:operation_instructions_configuration_protocol_unity}) will prompt you to add your StimulHeat device's public name (e.g., 0F6C), then click \textit{Configuration Complete}. On the second screen (see Figure~\ref{fig:operation_instructions_command_protocol_unity}), there is a panel for each of your connected StimulHeat device, from which you can switch the command type, i.e., Temperature or Flow, send commands and turn off the device.

\begin{figure}[ht!]
    \centering
    \includegraphics[width=0.5\textwidth]{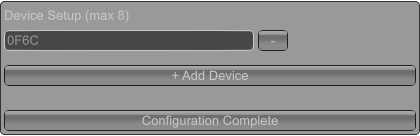}
    \caption{Configuration - Unity Desktop App (screenshot)}
    \label{fig:operation_instructions_configuration_protocol_unity}
\end{figure}

\begin{figure}[ht!]
    \centering
    \includegraphics[width=0.5\textwidth]{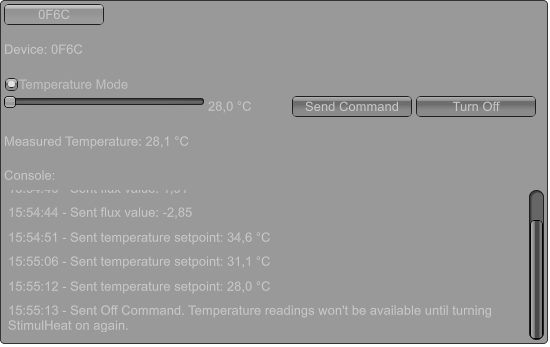}
    \caption{Control Panel - Unity Desktop App (screenshot)}
    \label{fig:operation_instructions_command_protocol_unity}
\end{figure}

\section{Validation and characterization}\label{Validation and characterization}

In order to illustrate the functioning of the StimulHeat system, two distinct experiments were conducted. The first study was purely technical, and its goal was to assess the performance of the hardware in terms of response time and stability when utilised on a particular subject. Thereafter, an additional experiment focused on user feedback was conducted, involving a number of participants (N=16) who were requested to give feedback of their thermal sensation perceived and offer their impressions following the testing of StimulHeat in a VE via a survey. The two protocols were formerly approbated by the research ethics committee of Université de Lyon’s ComUE under approval number 2023-09-21-00.

\subsection{\underline{Technical Characterization}}

\textbf{\textit{Materials and Methods:}}

\noindent The primary objective of the present study was to characterise Stimulheat in common-use conditions in VR. In order to achieve this goal, a single protocol was developed, which involved one healthy participant (male, aged 25). He was tasked to maintain the Valve Index controller with Stimulheat integration during a period of several technical trials. A total of 12 distinct thermal stimuli, 6 heat flow setpoints and 6 temperature setpoints, were administered in accordance with the subsequent protocol: firstly, a 5-second waiting period was maintained (\qty{0}{W} or \qty{31}{\celsius}, depending on whether the control was set to heat or temperature), followed by a 5-second thermal stimulus. It guaranteed the skin in contact with the TED to be returned to its initial state ($T_{skin}$ = \qty{31}{\celsius}) before conducting the subsequent stimulation. This temperature was measured in the specific context of the day it was recorded. The range of skin temperatures generally associated with a neutral thermal sensation is between \qty{30}{\celsius} and \qty{36}{\celsius}, according to the climatic context~\cite{Filingeri2017}. The lists of commands administered are detailed hereafter:

\begin{itemize}
\item{Heat flow : [\qty{-3}{W} ; \qty{-2}{W} ; \qty{-1}{W} ; \qty{1}{W} ; \qty{2}{W} ; \qty{3}{W}]}
\item{Temperature : [\qty{25}{\celsius} ; \qty{27}{\celsius} ; \qty{29}{\celsius} ; \qty{34}{\celsius} ; \qty{37}{\celsius} ; \qty{40}{\celsius}]}
\end{itemize} 

\noindent During the sequences, the temperature of the two thermistors was measured. The equation \ref{eq:Qa} was utilised to deduce the heat flow present in the TED, facilitating the comparison between the setpoint in heat flow and the real heat flow. Additionally, the temperature of the skin at the contact point between the palm and the TED was determined, permitting a comparison between the setpoint in temperature and actual temperature (see Figure \ref{fig:Technical-experience_Graphical_Performance of TED_Zoom}).

\noindent The objective of this technical characterisation is firstly to compare two methods of controlling the Stimulheat device and secondly to determine the advantages and disadvantages of each. This characterisation is made in use condition, and no consideration has been given to the thermal feelings perceived by the subject. The analysis of this feedback is part of the second study.

\textbf{\textit{Results and discussion:}}

\noindent The results are presented in Figure~\ref{fig:Technical Characterisation - Global Results}. First, it can be observed that heat control is more precise than temperature control. As outlined in the \hyperref[Hardware description]{2. Hardware description} section, it is directly related to the absence of a PID controller for heat control. Since a Peltier module is controlled via current, which directly influences the heat flow, the precision of heat control stems from this direct relationship.

\noindent Second, the response time of the control methods can be compared. For heat control, the response is instantaneous because the current value is derived using equation~(\ref{eq:Qa}). In contrast, with the current PID parameter values described in \hyperref[Hardware description]{2. Hardware description}, the system operates with a temperature change rate of approximately \qty{2.25}{\celsius/\second}. This allows us to conclude that heat control responds more quickly than temperature control. This result is expected, since a TED's heat flow is directly linked to the value of the current injected into it. In contrast, the temperature measured at the point of contact with the skin is not directly correlated with the injected current. This enhances the accuracy and robustness of the experimental conditions in heat flow-based control in comparison to temperature-based control, as it is independent of the PID parameter values.

\begin{figure}[ht!]
    \centering
    \begin{subfigure}[]{0.4\textwidth}
        \centering
        \includegraphics[width=1.2\linewidth]{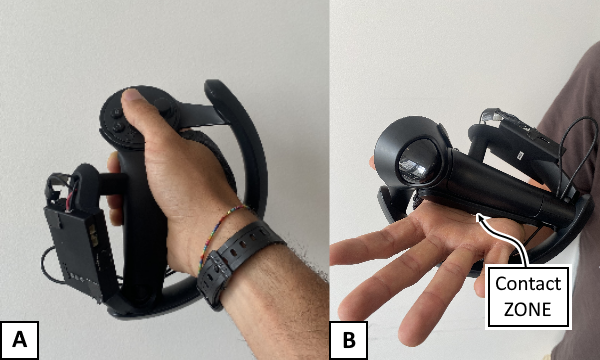}
        \caption{\textbf{A:} Overview of the characterization protocol - \textbf{B:} Close-up of the contact surface stimulated during the characterization protocol}
    \end{subfigure}
    \caption{Technical Characterisation - Protocol Picture}
    \label{fig:Technical-experience_Graphical_Performance of TED_Zoom}
\end{figure}

\begin{figure}[hb!]
    \centering
    \begin{subfigure}[]{0.75\textwidth}
        \centering
        \includegraphics[width=\linewidth]{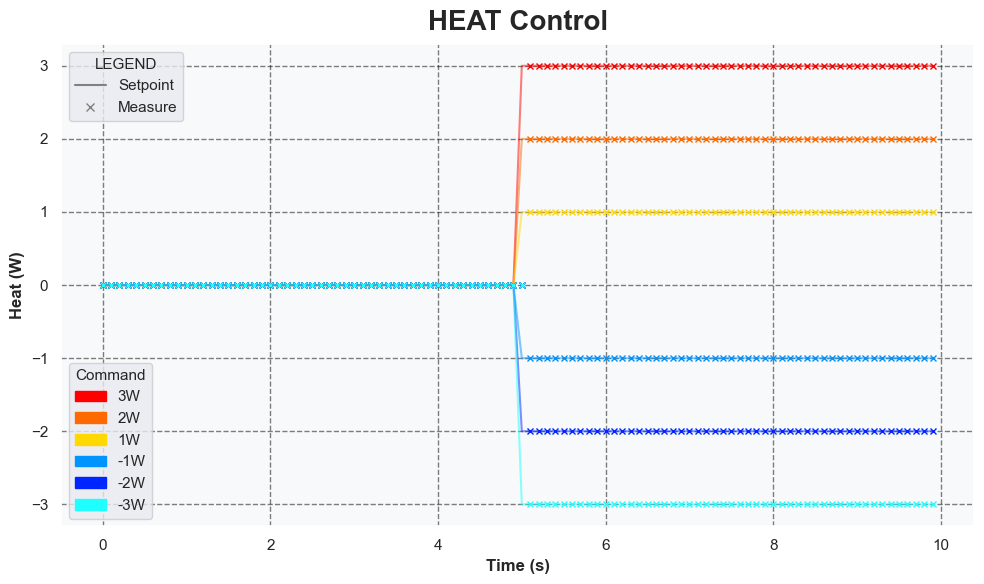}
    \end{subfigure}
    \begin{subfigure}[]{0.75\textwidth}
        \centering
        \includegraphics[width=\linewidth]{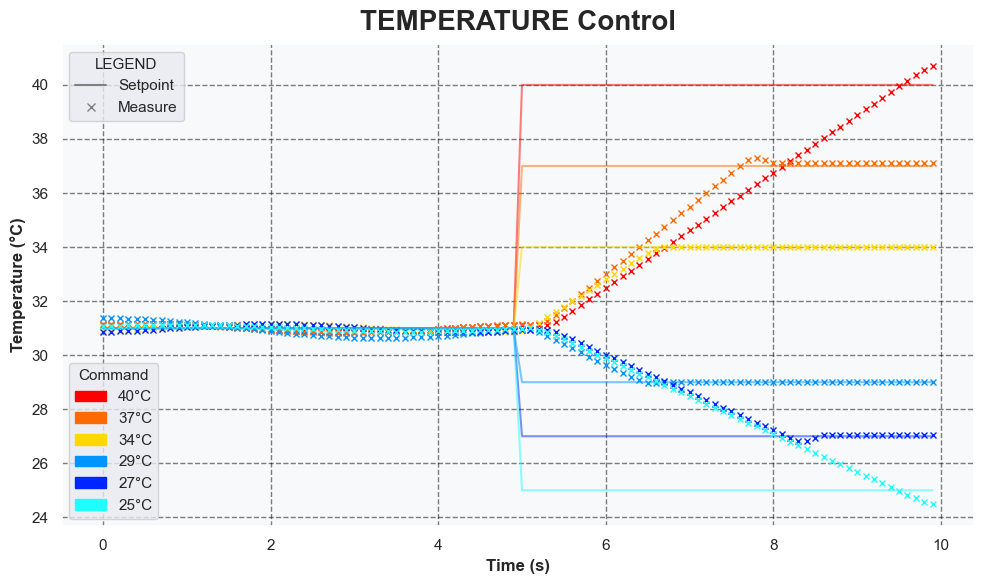}
    \end{subfigure}
    \caption{Technical Characterisation - Global Results}
    \label{fig:Technical Characterisation - Global Results}
\end{figure}

\subsection{\underline{User Validation}}

\textbf{\textit{Materials and Methods:}}

\noindent The goal of the second study was to validate the device in a VR setup use case. In this investigation, a novel protocol was developed, based on the subject's thermal sensation and the study of StimulHeat's user-friendliness and intrusiveness. The study was conducted with 16 participants (N = 16), ranging in age from \textbf{20 to 39}, including both male and female subjects (H = 9 ; F = 7). The participants were immersed in a VE for a period of approximately ten minutes and requested to grasp a blank totem on 20 occasions (2 for the tutorial and 18 for the experiment). Following a preliminary briefing, during which the explanation and consent document were given, the participants placed the VR headset and entered the VE. The visual representation of the VR scenario is illustrated in Figure \ref{fig:user_experience}. With each act of "picking up a totem" by the user, a heat command was applied to the StimulHeat controller on the hand which grasped the totem (either left or right, or twice if the user grabs the totem with both hands). The decision was taken to utilise the heat flow command due to its enhanced stability and reactivity in comparison to the temperature command. The command was either HOT (\qty{-2}{W}), NEUTRAL (\qty{0}{W}) or COLD (\qty{2}{W}). The thermal stimulation was administered for a duration of 5 seconds. In this period, the participants were requested to indicate the precise moment when they began to perceive the thermal stimuli by pressing a button on the Valve Index controller. This data contributes to our understanding of reaction time, which encompasses both the device's response time and the time required to perceive a sensory stimulus. At the end of the thermal stimulation, the totem then disappeared, and the participants had three seconds to consider the most accurate description of the perceived stimulation. They were then invited to select if the stimulus was perceived as either : HOT, NEUTRAL or COLD. The sequence of stimuli was administered identically to all participants and was as follows:

\begin{itemize}
\item{Heat flow : \newline\footnotesize[\qty{2}{W}~;~\qty{0}{W}~;~\qty{-2}{W}~;~\qty{2}{W}~;~\qty{-2}{W}~;~\qty{2}{W}~;~\qty{0}{W}~;~\qty{-2}{W}~;~\qty{0}{W}~;~\qty{-2}{W}~;~\qty{2}{W}~;~\qty{0}{W}~;~\qty{2}{W}~;~\qty{-2}{W}~;~\qty{2}{W}~;~\qty{0}{W}~;~\qty{-2}{W}~;~\qty{0}{W}]}
\end{itemize}

\noindent Following this, all the answers were collected and compared using a correlation matrix presented in figure \ref{fig:user-Thermal perception confusion matrix}, which plots the relationship between the command given and the thermal sensation perceived and communicated by the participants.

\begin{figure}[ht!]
    \centering
    \begin{subfigure}[]{0.95\textwidth}
       \centering
       \includegraphics[width=\linewidth]{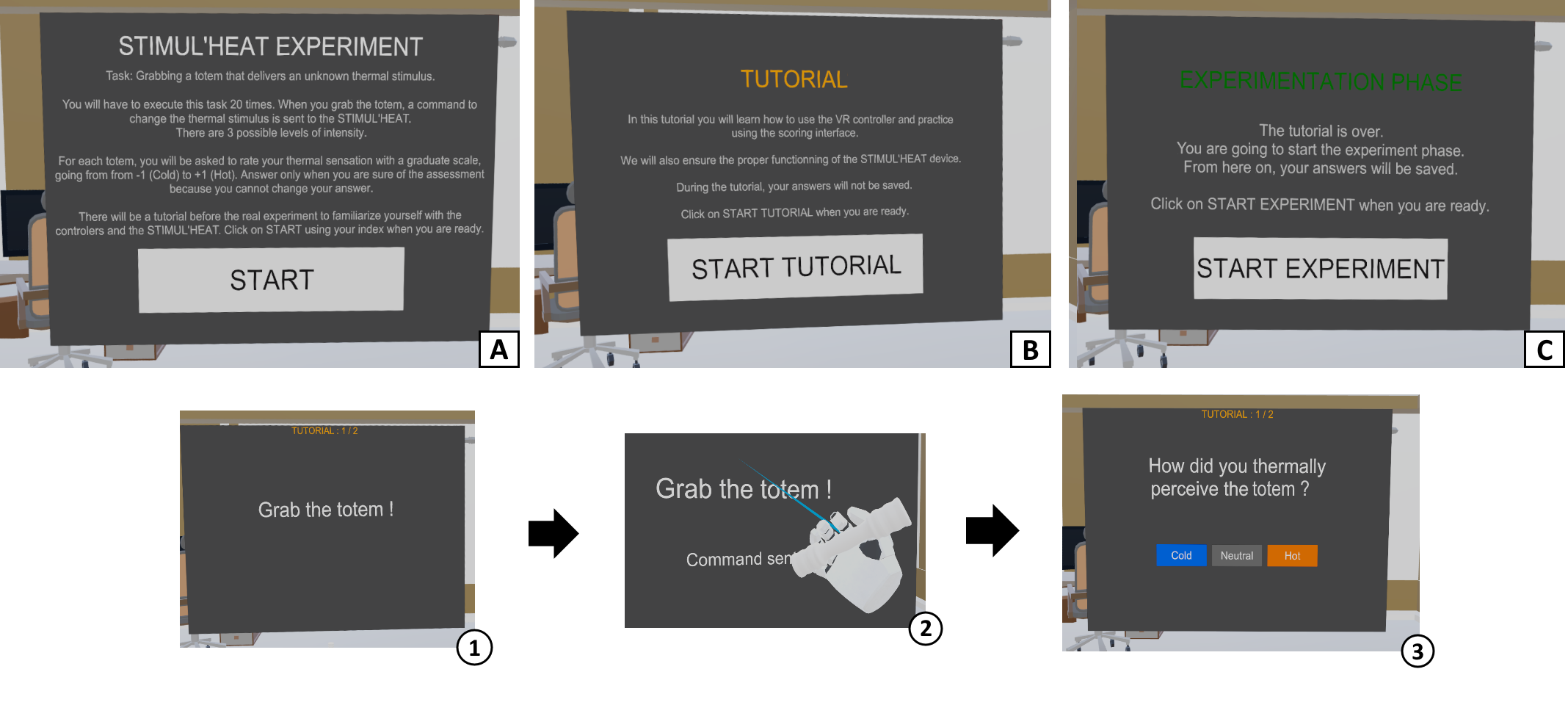}
       \caption{\textbf{Top:} VE interface shown to participants at each stage — \textbf{A}: Presentation panel at experiment start; \textbf{B}: Tutorial panel after clicking "Start"; \textbf{C}: Experimentation panel after tutorial completion. \textbf{Bottom:} Evaluation protocol for each totem — \textbf{1}: After a 5-second delay, the totem appears with a prompt to grab it; \textbf{2}: Upon grabbing, a 5-second heat stimulus is delivered; \textbf{3}: After a 3-second wait, participants rate the thermal sensation as COLD, NEUTRAL or HOT.}
    \end{subfigure}
    \caption{User experience - Protocol}
    \label{fig:user_experience}
    \vspace{0.5em}
\end{figure}

\noindent 
Comfort of use was also evaluated. To assess this, a survey with three affirmations was administered to each participant. We used a scale ranging from 1 to 7, where 1 denotes 'Strongly Disagree' and 7 denotes 'Strongly Agree' with the corresponding affirmation. In order to ensure an unbiased experience in relation to the use of VR, the panel of participants was composed of three different groups: one with people who have never used VR (Group "None"), one with participants who have already used VR at least once (Group "Beginner"), and one with participants who are familiar with the technology and have used it more than once (Group "Intermediate"). The number of participants in each group were $N_{None}$ = 5, $N_{Beginner}$ = 6, and $N_{Intermediate}$ = 5. The three affirmations of the survey are presented hereafter:

\begin{itemize}
\item{I experienced thermal discomfort on my hand while using the StimulHeat device during the experiment.}
\item{I experienced discomfort other than thermal while using the StimulHeat device during the experiment.}
\item{The controller felt too large for my hands and made it difficult to grasp objects.}
\end{itemize} 

\noindent The results obtained from the experiment are presented in Figure \ref{fig:user-Survey results group by level of VR experience}, thus enabling a comparative analysis of the responses to the affirmations according to each group.

\begin{figure}[ht!]
    \centering
    \includegraphics[width=0.5\textwidth]{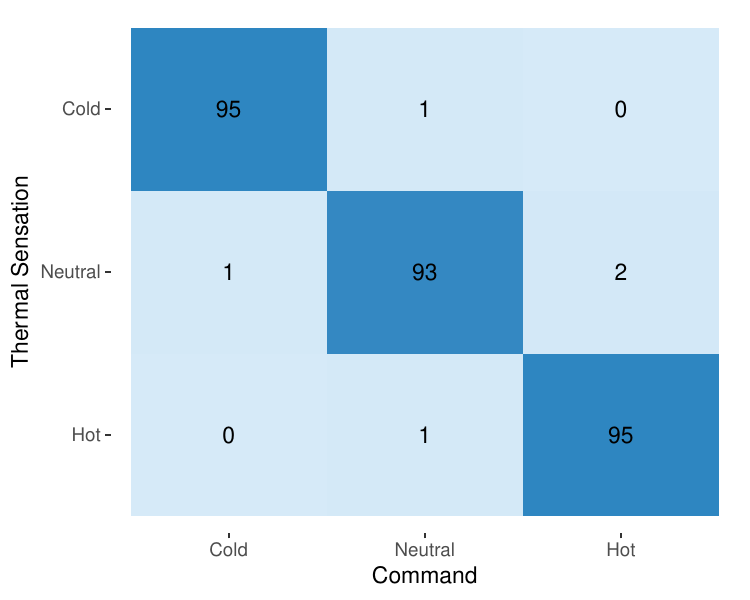}
    \caption{User validation – Thermal perception confusion matrix}
    \label{fig:user-Thermal perception confusion matrix}
\end{figure}

\begin{figure}[ht!]
    \centering
    \includegraphics[width=\textwidth]{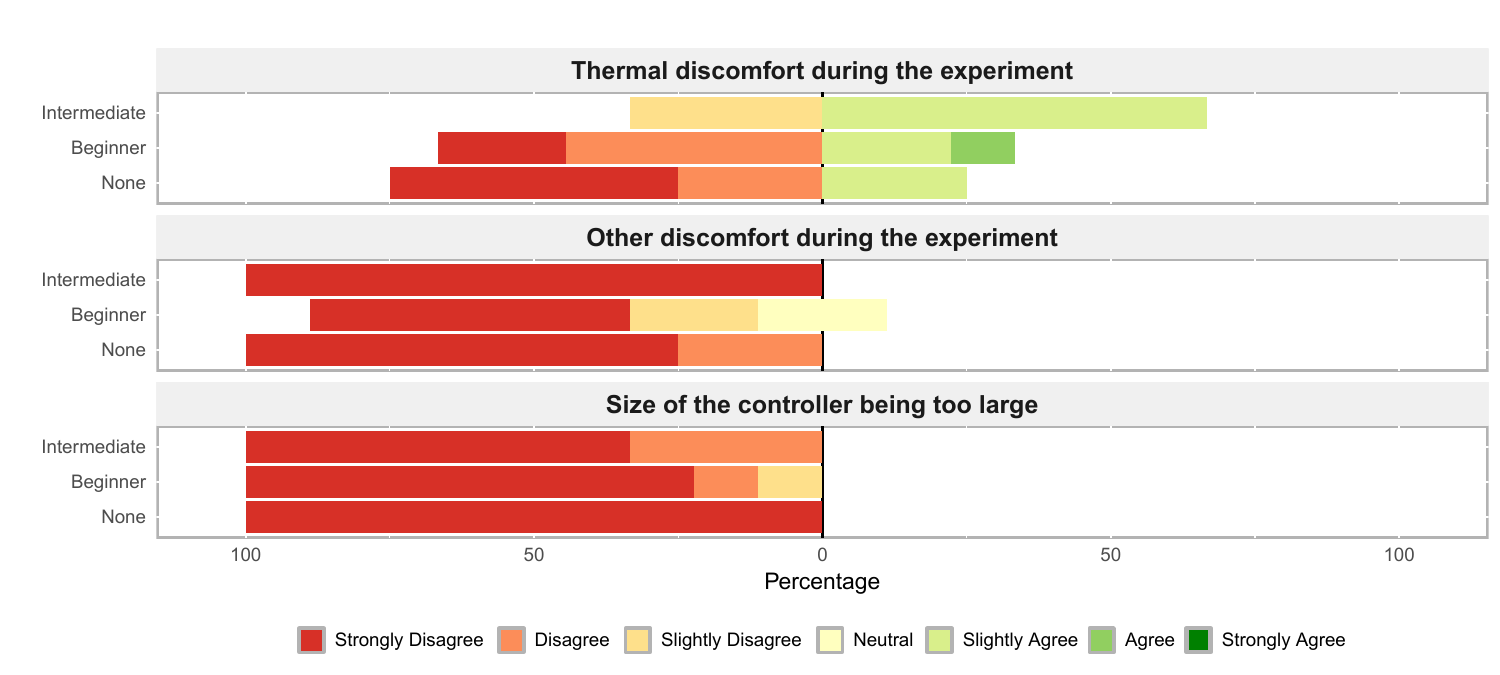}
    \caption{User validation – Survey results group by level of VR experience}
    \label{fig:user-Survey results group by level of VR experience}
\end{figure}

\textbf{\textit{Results and discussion:}}

\noindent Regarding the correlation matrix in the Figure \ref{fig:user-Thermal perception confusion matrix} which establishes a relationship between the perceived thermal sensation and the command, StimulHeat demonstrates a high degree of precision in regard to these three heat commands. Furthermore, when analysing the data exclusively in terms of failed predictions, it is notable that no instance of Hot Command has been experienced as Cold sensation, and vice versa. It can be added that in instances where the device is utilised to induce either a hot or a cold thermal sensation, there were two cases on 192 whose the command failed to generate such a sensation. Three cases on 96 were registered in which a thermal sensation was perceived, despite the absence of any command injection in the TED.

\noindent Next, regarding the boxplots in Figure \ref{fig:user-Survey results group by level of VR experience}, several deductions can be made. Regarding the initial statement on thermal discomfort, it appears that the intermediate group has encountered a higher level of discomfort in comparison to the beginner and none groups. This could possibly be explained by the fact that for the two groups mentioned above, this experience stimulates their curiosity and the discovery of the technology and perhaps inhibits their ability to feel the surrounding context. Nevertheless, the identified thermal discomfort remains low even for the Intermediate group's VR experience. Turning to the boxplot relating to the controller size perceived as excessive for the hand, it can be deduced from the responses that the incorporation of StimulHeat into the controller does not compromise the user experience or comfort during a VR session, as none of the participants from all groups expressed any opinion in favour of this statement. These results are supported by the boxplot associated with the second affirmation, which concerns the discomfort experienced during VR (virtual reality) sessions, different from thermal discomfort. The results of the study indicated that the majority of participants did not agree with the proposal, and thus that the integration design of the StimulHeat did not cause discomfort.

\section{Limitations and Conclusion}\label{Limitations and Conclusion}

\noindent One limitation of the system is it capacity to maintain low operating temperatures over extended periods. This issue is common in devices that employ TED when the associated heat dissipation solution is insufficiently effective. In particular, when StimulHeat is utilised in an extreme scenario for a period exceeding five minutes, such as when the stimulus is set to an extremely low temperature, the TED begins to lose its cooling capacity and may consequently start to heat up instead. As a result, the system's performance degrades, and the user may perceive a COLD or even NEUTRAL sensation on the contact surface instead of the expected VERY COLD stimulus. To mitigate this limitation, we used an efficient TED and applied a proper control strategy by using a DC power source instead of a PWM-controlled output current, which is more suitable for Peltier elements. Despite these technical choices, it remains challenging to use StimulHeat in VERY COLD scenarios over extended durations. Therefore, it is necessary to anticipate and prevent this specific use case. In the context of virtual reality, this can be achieved for instance by automatically disabling the VERY COLD command before reaching the system's thermal threshold, for example, after three minutes of continuous operation. 
Additionally, a technical alternative solution exists, which involves maintaining a constant temperature on the external surface of the TED for $T_e$ (see Figure \ref{fig:peltier_schematic}). It can be demonstrated that, by maintaining a temperature of \qty{30}{\celsius} on the hot side, the injection of the current will only result in a change to the temperature of the absorbed side without modifying the one on the emitted face. In principle, this would enable the utilisation of StimulHeat in extreme use cases. However, this would come at the cost of significantly higher energy consumption, increased weight, and reduced user comfort due to greater intrusiveness. Our research team is currently investigating the aforementioned issue. Currently, only a ceramic heat sink (MPC202025T) passively helps to maintain this temperature. In the future, additional cooling solutions, such as a micro water-cooling system, could be implemented to better achieve this objective, as has been successfully achieved by other researchers~\cite{Itao2016}.

\noindent In conclusion, this paper presents a solution for haptic thermal feedback in VR called StimulHeat. This solution has been designed for Valve Index controllers, but is also adaptable to other controllers by redesigning the shell. The device is equipped with two control systems. The first system functions through temperature regulation, whilst the second utilises a heat flow command, which is more efficient and direct method to drive the TED. Moreover, it has been engineered to operate with minimal power consumption, with a maximum power rating of \qty{2.22}{W}, and a battery life of several hours. The device can be remotely controlled via BLE (Bluetooth Low Energy) communication. In order to validate the system, both objective and subjective characterization studies were conducted. Our results show that the device demonstrates a high degree of efficacy in creating thermal sensations. Furthermore, the design is perceived as comfortable by the majority of participants. 


\textbf{Ethics statements}

\noindent As the study incorporates human subjects, consent was obtained from the volunteers who used StimulHeat via a brief informational session prior to the experiment and a consent agreement. This study was approved by the research ethics committee of Université de Lyon’s ComUE, under approval number 2023-09-21-005.\\

\noindent
\textbf{CRediT author statement}

\noindent \textbf{Matthieu Mesnage}: Conceptualization, Methodology, Software, Validation, Writing, Review, Editing}\\
\textbf{Sophie Villenave}: Formal analysis, Data Curation, Visualization, Software, Validation, Review, Editing\\
\textbf{Bertrand Massot}: Conceptualization, Software, Validation, Writing, Review Editing, Supervision\\
\textbf{Matthieu Blanchard}: Software, Validation, Review, Editing\\
\textbf{Pierre Raimbaud}: Review, Editing, Supervision\\
\textbf{Guillaume Lavoué}: Software, Validation, Review, Editing, Supervision\\
\textbf{Claudine Gehin}: Supervision\\

\noindent
\textbf{Acknowledgements}

\noindent This work was supported in part by the French National Reasearch Agency (ANR) under the grant ANR-22-CE31-0023-03 RENFORCE.

\bibliography{hardwareX}

\end{document}